\documentclass[useAMS,usenatbib]{mn2e}

\usepackage{amssymb}
\usepackage{amsfonts}
\usepackage{natbib}
\usepackage{epsfig}
\usepackage{mncite}

\newcommand{\de}{\mbox{d}}

\begin{document}

\title[Seed black holes]{The mass function of high redshift seed black
holes}

\author[Lodato \&  Natarajan] {Giuseppe Lodato$^{1,2}$ and Priyamvada 
Natarajan$^{3,4}$\\
$^1$ Institute of Astronomy, Madingley Road, Cambridge, CB3 0HA\\
$^2$ Department of Physics and Astronomy, University of Leicester, Leicester,
LE1 7RH\\
$^3$ Department of Astronomy, Yale University, P. O. Box 208101, 
New Haven, CT 06511-208101, USA \\
$^4$ Department of Physics, Yale University, P. O. Box 208120, 
New Haven, CT 06520-208120, USA}

\maketitle

\begin{abstract}
  In this paper we derive the mass function of seed black holes that result
  from the central mass concentrated via disc accretion in collapsed haloes at
  redshift $z\approx 15$.  Using standard arguments including stability, we
  show that these pre-galactic discs can assemble a significant mass
  concentration in the inner regions, providing fuel for the formation and
  initial growth of super-massive black holes.  Assuming that these mass
  concentrations do result in central seed black holes, we determine the mass
  distribution of these seeds as a function of key halo properties.  The seed
  mass distribution determined here turns out to be asymmetric and skewed to
  higher masses.  Starting with these initial seeds, building up to $10^9$
  solar masses by $z = 6$ to power the bright quasars is not a problem in the
  standard LCDM cosmogony. These seed black holes in gas rich environments are
  likely to grow into the supermassive black holes at later times via mergers
  and accretion.  Gas accretion onto these seeds at high redshift will produce
  miniquasars that likely play an important role in the reionization of the
  Universe. Some of these seed black holes on the other hand could be
  wandering in galaxy haloes as a consequence of frequent mergers, powering
  the off-nuclear ultra-luminous X-ray sources detected in nearby galaxies.
 \end{abstract}


\section{Introduction}

It is becoming increasingly clear that supermasssive black holes assemble in
galactic nuclei, and that their growth is in tandem with that of the stellar
component \citep{merritt01,tremaine02}.  Optically bright quasars powered by
accretion onto black holes are now detected out to redshifts of $z > 6$ when
the Universe was barely 7\% of its current age \citep{fan04,fan06}.  The
luminosities of these high redshift quasars imply black hole masses $M_{\rm
  BH} > 10^9\,M_{\odot}$. Assembling these large black hole masses by this
early epoch starting from remnants of the first generation of metal free stars
has been a challenge for models.  Models that describe the growth and
accretion history of supermassive black holes commence with high-redshift
seeds derived from Pop-III stars \citep{haiman98,volonteri05,wyithe05}.  Some
suggestions to accomplish this invoke super-Eddington accretion rates for
brief periods of time \citep{volonteri05b}. Alternatively, it has been
suggested that the formation of more massive seeds ab-initio
\citep{eisenstein95,koushiappas04}, possibly through the collapse of a massive
quasi-star \citep{begelman06}, could also alleviate the problem of building up
supermassive black hole masses to the required values by $z = 6$. We explore
this possibility further in this paper.

In a previous paper (\citealt{LN06}, hereafter LN), we have discussed a model
in which supermassive black holes form and grow in the center of dark matter
haloes directly from the gas content of the halo, which is channelled to the
center by the torques arising from gravitational instabilities that develop in
assembling massive pre-galactic discs. Models similar to ours, but using
different sets of arguments were also proposed in the past
\citep{eisenstein95,haehnelt93,koushiappas04,begelman06}. In particular, our
model improves upon a similar proposal by \citet{koushiappas04} (hereafter
KBD) by self-consistently considering the evolution of the disc surface
density during the accretion process and by taking into account the
possibility that a very massive disc might undergo fragmentation and form
stars rather than exclusively transport gas to the center. In this way, we
obtain a robust determination of the amount of mass accreted in the center of
the pre-galactic disc, which is remarkably independent of many parameters,
like the angular momentum distribution within the halo, the details of the
viscosity (as long as it is driven by gravitational instabilities) and the
viscous timescale of the disc.  As described further below, the accreted mass
only depends on halo structural parameters (like its mass $M$, the spin
parameter $\lambda$ and the fraction of baryonic mass that collapses to a disc
$m_{\rm d}$), and on some well constrained gas dynamical parameters, like the
critical value of the Toomre parameter for the onset of instability $Q_{\rm
  c}\approx 1-2$, and the gas temperature $T_{\rm gas}$ ($\approx 4000$K for
atomic hydrogen cooling, and $\approx 300$K for molecular hydrogen cooling).

In LN we were mostly concerned with the details of the hydrodynamics of the
disc. Here we further develop our model by calculating the resulting mass
function of such central mass concentrations, available for the growth of seed
black holes, and their volume density as a function of redshift, with the
standard prescription for the halo mass function \citep{sheth99}.  This will
allow us to further clarify the role of the different ingredients (accretion,
fragmentation) of the model in the determination of the black hole population.
The paper is organized as follows: in section 2 we summarize the main features
of our model, in section 3 we calculate the mass function and density as a
function of redshift, in section 4 we discuss our results and potential
observational signatures.

\section{The model}

In the model described in greater detail in an earlier paper (LN), we consider
a dark matter halo of mass $M$ and virial temperature $T_{\rm vir}$ (for
simplicity, the density profile is assumed to be that of an isothermal
sphere), containing gas which we assume to be of primordial composition, i.e.
not enriched by metals, for which the cooling function is dominated by
hydrogen. In the following, when numbers are needed, we will assume that the
fraction of mass that collapses and forms a disc is $M_{\rm gas}=m_{\rm d}M$,
with $m_{\rm d}=0.05$ \citep{mo98}.

The angular momentum of the dark matter halo $J$ is expressed in terms of its
spin parameter $\lambda=J|E|^{1/2}/GM^{5/2}$, where $E$ is its total energy.
The probability distribution of the spin parameter of dark matter haloes can
be obtained from cosmological N-body simulations \citep{warren92} and is well
described by:
\begin{equation}
p(\lambda)\de \lambda=\frac{1}{\sqrt{2\pi}\sigma_{\lambda}}
\exp\left[-\frac{\ln^2(\lambda/\bar{\lambda})}{2\sigma_{\lambda}}\right]
\frac{\de\lambda}{\lambda},
\end{equation}
where $\bar{\lambda}=0.05$ and $\sigma_{\lambda}=0.5$. We further
assume that the angular momentum of the baryonic component is a fraction
$j_{\rm d}=m_{\rm d}$ of the halo angular momentum $J$.

If the virial temperature of the halo $T_{\rm vir}>T_{\rm gas}$, the
gas collapses and forms a rotationally supported disc, with circular
velocity $V_{\rm h}$, determined by the gravitational field of the
halo.
For low values of the spin parameter $\lambda$ the resulting disc can
be compact and dense and is subject to gravitational
instabilities. This occurs when the stability parameter $Q$:
\begin{equation}
Q=\frac{c_{\rm s}\kappa}{\pi G \Sigma}=\sqrt{2}\frac{c_{\rm s}V_{\rm
h}}{\pi G\Sigma R}\lesssim Q_{\rm c},
\label{Q}
\end{equation}
where $c_{\rm s}$ is the sound speed, $\kappa=\sqrt{2}V_{\rm h}/R$ is
the epicyclic frequency, $R$ is the cylindrical radial coordinate, and
$\Sigma$ is the surface mass density and where the critical value
$Q_{\rm c}\approx 1-2$.

The subsequent evolution of the gravitationally unstable disc is determined by
the competition between two physical processes: accretion and fragmentation.
Indeed, gravitational instabilities lead to the development of
non-axisymmetric spiral structures (or a bar), which lead to an effective
redistribution of angular momentum, thus possibly feeding a growing seed black
hole in the center. This process stops when the amount of mass transported in
the center, that we refer to suggestively as $M_{\rm BH}$, is enough to make
the disc marginally stable. This can be computed easily from the stability
criterion in Eq. (\ref{Q}) and from the disc properties, determined from the
halo mass and angular momentum \citep{mo98}. In this way we obtain that the
mass accumulated in the center of the halo is given by:
\begin{equation}
M_{\rm BH}= \left\{\begin{array}{ll}

\displaystyle m_{\rm d}M\left[1-\sqrt{\frac{8\lambda}{m_{\rm d}Q_{\rm c}}
\left(\frac{j_{\rm d}}{m_{\rm d}}\right)\left(\frac{T_{\rm gas}}{T_{\rm
vir}}\right)^{1/2}}\right] & \lambda<\lambda_{\rm max} \\
0 & \lambda>\lambda_{\rm max}
\end{array}
\right.
\label{mbh}
\end{equation}
where $\lambda_{\rm max}=m_{\rm d}Q_{\rm c}/8(m_{\rm d}/j_{\rm d}) (T_{\rm
  vir}/T_{\rm gas})^{1/2} $ is the maximum halo spin parameter for which the
disc is gravitationally unstable.

\begin{figure}
\centerline{\epsfig{figure=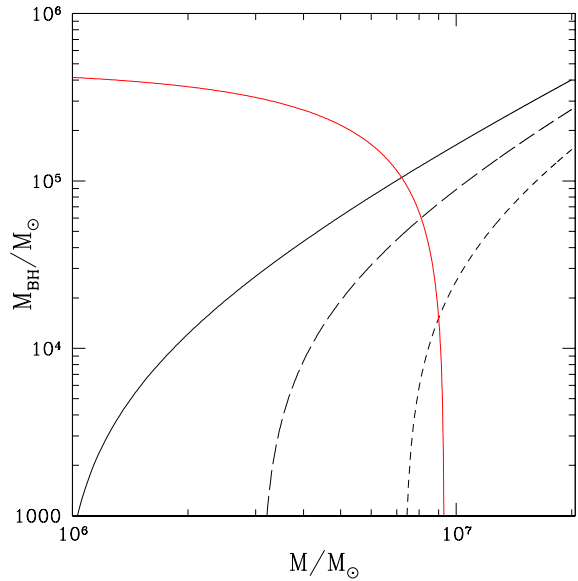,width=0.45\textwidth}}            
\caption{\small Mass available for the growth of a seed black hole as a
  function of halo mass, as obtained from Equation (\ref{mbh}). The plots
  refer to the following choice of parameters: $Q_{\rm c}=2$, $T_{\rm
    gas}=4000$K, $m_{\rm d}=j_{\rm d}=0.05$, $\lambda = 0.01$ (solid line),
  $\lambda=0.015$ (long-dashed line), $\lambda=0.02$ (short-dashed line). The
  red curves shows the threshold for fragmentation from Equation (\ref{frag}),
  with $\alpha_{\rm c}=0.06$. Haloes on the right of the red line give rise to
  fragmenting discs in our model.}
\label{fig:mass}
\end{figure}

It is worth noting here that the estimates obtained using Eq. (\ref{mbh}) are
in fact  upper limits to the actual black hole mass, since a fraction of the
infalling gas mass might not be accepted by the growing black hole, if, for
example, the rate at which the mass is delivered is significantly
super-Eddington.  This possibility is discussed further below.

However, for large halo mass, the internal torques needed to
redistribute the excess baryonic mass become too large to be
sustained by the disc, which then undergoes fragmentation. This occurs
when the virial temperature exceeds a critical value $T_{\rm
max}$, given by:
\begin{equation}
\frac{T_{\rm max}}{T_{\rm gas}}>\left(\frac{4\alpha_{\rm c}}{m_{\rm
d}}\frac{1}{1+M_{\rm BH}/m_{\rm d}M}\right)^{2/3},
\label{frag}
\end{equation}
where $\alpha_{\rm c}\approx 0.06$ is a dimensionless parameter measuring the
critical gravitational torque above which the disc fragments \citep{RLA05}
(note that this value has been determined for thin, Keplerian discs, and might
be slightly different in the case considered here, see also LN). It is not
easy to estimate how much mass can still be accreted once the fragmentation
threshold is overcome. The fragmentation process is fast, occuring on the
local dynamical timescale, and will stop once enough mass has been converted
into stars to make the disc stable (i.e. the mass converted into stars will be
of the order of $M_{\rm BH}$, neglecting possible feedback effects), hence
preventing further accretion, in the absence of other angular momentum removal
mechanism. A conservative assumption is that, when fragmentation occurs, no
mass is transported to the center. In the following we will discuss the
resulting mass function for the two extreme cases where fragmentation
completely quenches accretion and where it is not taken into account.

Fig. \ref{fig:mass} illustrates the relationship between halo mass and
potential black hole mass in our model, based on Equations (\ref{mbh}) and
(\ref{frag}).

One of the main ingredients in our model is the gas temperature $T_{\rm
gas}$. This is computed in LN by assuming that the gas is in thermal
equilibrium where radiative losses are balanced by the heating provided
by the accretion process. We find that if the gas is atomic hydrogen
$T_{\rm gas}\approx 4000$K. If the gas is instead molecular hydrogen,
the disc is much colder with $T_{\rm gas}\approx 300$K. 

\section{Results} 

\begin{figure}
\centerline{\epsfig{figure=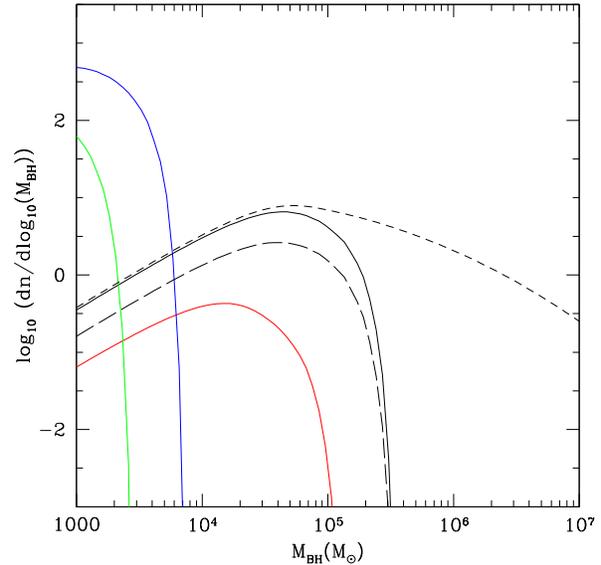,width=0.45\textwidth}}            
\caption{\small Mass function of high redshift black holes. The black solid
line refers to the following choice of parameters: $z=10$, $T_{\rm
gas}=4000$K and $Q_{\rm c}=2$. The short-dashed line illustrates the
effect of removing the threshold for disc fragmentation. The
long-dashed line shows the effect of decreasing th critical value of
$Q$ to 1.5. The red line shows the results for $Q_{\rm c}=2$ at
$z=20$. The two remaining lines refer to the cold case where $T_{\rm
gas}=300$K at $z=10$ (blue line) and at $z=20$ (green line).}
\label{fig:bhmf}
\end{figure}

\begin{figure}
\centerline{\epsfig{figure=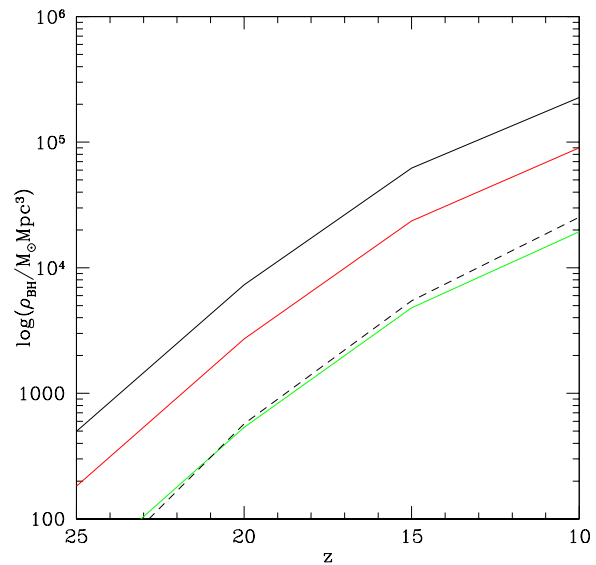,width=0.45\textwidth}}            
\caption{\small Integrated mass density of black holes as a function of
redshift. The three solid lines refer to different choices of $Q_{\rm
c}$ for the `hot' $T_{\rm gas}=4000$K model: $Q_{\rm c}=3$ (black),
$Q_{\rm c}=2$ (red) and $Q_{\rm c}=1$ (green). The dashed line refers
to $T_{\rm gas}=300$K and $Q_{\rm c}=1$.}
\label{fig:dens}
\end{figure}

\subsection{The mass function of seed black holes}

From the model outlined in the previous section, using the mass function of
dark matter haloes computed in the LCDM model of structure formation, we can
easily obtain the mass function of the central mass concentration at high
redshifts. Indeed, given the halo mass $M$, the probability of accumulating at
the center of the disc a mass between $M_{\rm BH}$ and $M_{\rm BH}+\mbox{d}
M_{\rm BH}$ is given by:
\begin{equation}
P(M_{\rm BH})\mbox{d}M_{\rm BH}=p(\lambda(M,M_{\rm
BH}))\mbox{d}\lambda,
\label{prob}
\end{equation}
where $\lambda(M,M_{\rm BH})$ is obtained inverting Eq. (\ref{mbh}). To obtain
the mass function at a given redshift we need to integrate the probability
given in Eq. (\ref{prob}) over the halo mass function at that redshift,
$\mbox{d} n/\mbox{d} M$:
\begin{eqnarray}
& \nonumber \displaystyle \frac{\mbox{d}n}{\mbox{d}M_{\rm BH}}(M_{\rm
 BH}; z) & = \\
& \displaystyle \int_{M(T_{\rm
min})} ^{M(T_{\rm
max})} \frac{\mbox{d}n}{\mbox{d}M}(M;z) & p[\lambda (M_{\rm BH},M)]
\left|\frac{\mbox{d}\lambda}{\mbox{d}M_{\rm BH}}\right|\mbox{d} M,
\label{bhmf}
\end{eqnarray}
where the integration limits are obtained from the constraints that the
halo virial temperature has to be larger than $T_{\rm min}=T_{\rm gas}$
(in order for the gas to collapse and form a disc) and has to be
smaller than $T_{\rm max}$, obtained from Eq. (\ref{frag}), in order to
prevent the disc from fragmenting.

For the input halo mass function we adopt the standard distribution based on
the Press-Schechter formalism \citep{sheth99} and a LCDM cosmology with the
following characteristic parameters: $\Omega_{\rm m}=0.24$,
$\Omega_{\Lambda}=0.76$ and $\sigma_8=0.74$.  The resulting mass functions are
plotted in Fig. \ref{fig:bhmf}.  The solid line refers to the case where the
gas temperature is $T_{\rm gas}=4000$K, the critical value of $Q$ is $Q_{\rm
  c}=2$, computed at redshift $z=10$. The short-dashed line illustrates the
effects of not including the fragmentation criterion, i.e.  computed for the
same parameters but by removing the upper limit on the halo mass in the
integration in Eq. (\ref{bhmf}). It can be seen that fragmentation removes the
highest masses from the population and gives an effective cut-off at a mass
$\approx 2-3~10^{5}M_{\odot}$. The long-dashed line illustrates the effect of
decreasing the critical value of the stability parameter to $Q_{\rm c}=1.5$.
In this case the discs are more stable against gravitational instabilities and
the effect is simply to reduce the overall normalization of the mass function.
The red line shows the result for the same parameters as the black one, but at
a higher redshift $z=20$. As expected, at higher redshifts the typical masses
obtained from our model are scaled down due to lower value of the typical host
dark matter halo mass. The green and the blue lines refer to the case where
the gas is assumed to cool via molecular hydrogen to $T_{\rm gas}=300$K, at
$z=20$ (green line) and at $z=10$ (blue line). As already noted in LN, for the
cold case the typical central concentrations have masses of the order of
$10^3M_{\odot}$ and are more abundant, reflecting the larger abundance of low
mass haloes.

\subsection{Integrated BH mass density}

The integrated mass density of the central concentration as a function of
redshift can be easily obtained from the mass function:

\begin{equation}
\rho_{\rm BH}(z) = \int\frac{\mbox{d} n}{\mbox{d} M_{\rm BH}}(M_{\rm
BH},z) M_{\rm
BH} \mbox{d}M_{\rm BH}.
\end{equation}
The results are shown in Fig. \ref{fig:dens}. The three solid lines refer to
the `hot' model with cooling dominated by atomic hydrogen and $T_{\rm
  gas}=4000$K. The value of the critical $Q$ for gravitational instabilities
is $Q_{\rm c}=3$ (black line), $Q_{\rm c}=2$ (red line), and $Q_{\rm c}=1$
(green line). As expected, decreasing the value of $Q_{\rm c}$ makes the discs
relatively more stable so that the corresponding black hole density decreases.
The dashed line refers to the `cold' case, dominated by molecular hydrogen
cooling and with $T_{\rm gas}=300$K and $Q_{\rm c}=1$. Interestingly, the
integrated black hole density in this case (where smaller central mass
concentrations are produced) is comparable to the `hot' case. This is due to
the fact that in the cold case the population is dominated by a large number
of small ($M_{\rm BH}\approx 10^3M_{\odot}$) objects in the relatively more
numerous small haloes, which are absent from the `hot' case, because the gas
in such small haloes has a temperature larger than the virial temperature and
does not collapse to form a disc.

The present day black hole density, as derived by Soltan's argument is between
$2-4~10^5M_{\odot}/\mbox{Mpc}^3$ \citep{yu02}.  It would then naively seem
that our model might overestimate this quantity at low redshifts. It should be
however kept in mind that our proposed model strictly offers a quantitative
estimate of the black hole mass function only at high redshift, before the
intergalactic medium is enriched by metals. It is then interesting to see that
our model is able to provide a reasonable black hole density of a few times
$10^4 M_{\odot}/\mbox{Mpc}^3$ at a redshift, say, $z=15$ to allow for further
AGN-like growth to the present day density. The consequences of using the seed
mass function derived above to compute the supermassive black hole mass
function at low redshift using the merger-tree approach is currently underway
(Volonteri, Lodato \& Natarajan, in preparation).

Another important aspect to be taken into consideration is when the transition
occurs from mostly molecular gas to mostly atomic gas. It is reasonable to
assume that at very high redshift the gas is predominantly molecular becoming
atomic after the birth of the first stars, when the UV background has
increased substantially \citep{oh03}.  A detailed investigation of this
problem is however, beyond the scope of the present paper.

\begin{figure}
\centerline{\epsfig{figure=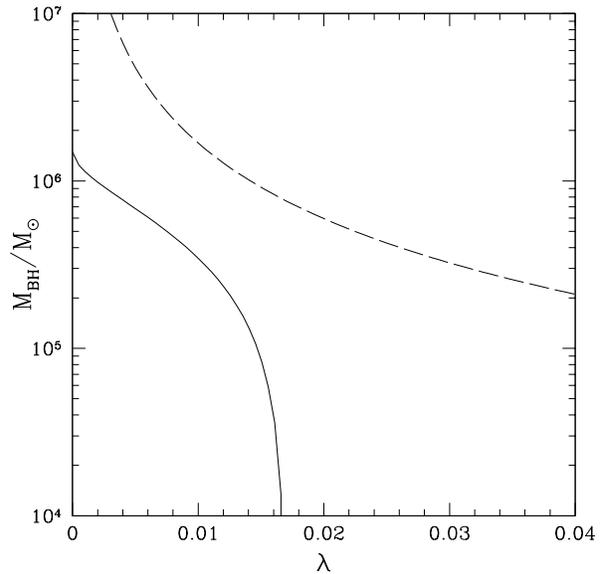,width=0.45\textwidth}}            
\caption{\small The relationship between spin parameter and black hole mass
  for a halo of mass $M=3~10^7M_{\odot}$ at $z=14$, according to the LN model
  (solid line) and to the KBD model (dashed line). For the KBD model, the
  additional parameter $t_{\rm visc}$ was assumed to be 10Myrs, while for the
  LN model we have assumed $Q_{\rm c}=2$.}
\label{fig:kvsln}
\end{figure}

\section{Comparison to earlier work} 

A model similar to our own has been also discussed by KBD. A detailed
discussion of similarities and differences between the two models is provided
in LN. Here we point out how these differences affect the resulting mass
function. There are two main differences between the two models: {\it (i)} we
include the evolution of the disc surface density due to the redistribution of
angular momentum caused by gravitational instabilities, while KBD only
consider the evolution induced by the infall of matter onto the disc. {\it
  (ii)} We include the effects of disc fragmentation, which are neglected by
KBD. As a result of point {\it (i)} above, the KBD results depend on some
additional free parameters which are implicitly (and self-consistently)
included in our model, like the viscous time-scale in the disc. These affect
the functional relationship between central mass and spin parameter for a
given halo mass dramatically.  In Fig.  \ref{fig:kvsln} we plot $M_{\rm BH}$
as a function of $\lambda$ for a given halo mass according to our model and
the corresponding plot obtained using the KBD model (see Figure caption for
details). We note that for the same structural parameters our model predicts
lower masses than KBD.

The relationship between $M_{\rm BH}$ and $\lambda$ according to KBD is a
simple power-law, so that for a given halo mass, the black hole mass
distribution is essentially log-normal, as the spin parameter distribution,
while this is not the case in our model. Our results suggest that dark matter
haloes with high spin values (above a certain value of the spin parameter
$\lambda$) do not host black holes at early times. This is because for a given
$M$, if the disc is not compact enough (or conversely, for a given $\lambda$,
if the disc is not massive enough) it is stable to gravitational
instabilities. Such haloes likely have more conventional seeds from remnants
of the first generation of stars or eventually host black holes as a
consequence of mergers at later times \citep{narayanan00}.

Additionally, the black hole mass function as derived from the LN model is
also skewed to high black hole masses, due to the effective mass cut-off set
by the onset of fragmentation in very massive discs, as shown in
Fig. \ref{fig:bhmf}.

\section{Conclusions and Discussion}

One of the key assumptions made in this work is that the centrally
concentrated mass in pre-galactic discs eventually forms a black hole.  As
mentioned above in Section~2, not all of the mass that flows into the center
of the halo might be accepted by a growing black hole, so that our mass
estimates should, strictly speaking, be considered upper limits. We wish to
stress that in this paper we are mostly concerned with determining the total
mass budget of such systems, rather than in the eventual fate of the accreting
matter.  Detailed models of the ultimate fate of the gas as it forms a seed
black hole have shown how the growth can be super-Eddington
\citep{volonteri05b,begelman06}. Alternatively, the accretion flow at the
smallest scale might break down and become chaotic. In such a situation
\citep{king06} the accretion process does not spin up the hole efficiently and
the accretion efficiency is thus kept low, to $\epsilon\approx 0.05$ (see also
\citealt{LP06}), effectively increasing the Eddington accretion rate.

Once a seed black hole with a mass of a few times $10^4-10^5 M_{\odot}$ has
formed (cf. the mass function shown in Fig. \ref{fig:bhmf} at $z=20$), it can
initially keep accreting following our model. A strict upper limit to the
accretion rates through the disc in our model is given by the threshold for
fragmentation, which is $\approx 0.01 M_{\odot}/$yr (see LN), which
corresponds to the Eddington limit for a $10^5M_{\odot}$ black hole, with an
efficiency of $\epsilon = 0.05$ \citep{king06}. Therefore, we can be confident
that in the majority of cases most of the available mass will indeed be
accepted by the growing black hole, so that our upper limits for the black
hole density are fairly realistic estimates of the actual values.

Our simplified analytical model for accretion essentially considers isolated
haloes, while the Press-Shechter formalism adopted to determine the mass
function is based on the merger history of the dark matter haloes. Mergers can
have two effects on the results described here. Firstly, a merger might
disrupt the disc before it has enough time to form and accrete to the center.
These two processes occur on the same timescale, i.e. the internal dynamical
time of the halo, which is generally shorter than the survival time of the
halo (see \citealt{lacey93}, Fig. 5), so our model is consistent from this
point of view. A second point is that a halo at a given redshift might already
contain a hole formed previously in the parent haloes. However, as long as the
growth of the seed is dominated by accretion rather than by merger of
different seeds (which is generally the case, see \citealt{volonteri03}) we
can still use our formalism to determine the new seed mass. At a given
redshift, some seed black holes will have formed anew from the lowest mass
haloes, while some others (contained in the most massive haloes) will already
have formed earlier but they will still have grown to a new, much larger mass,
that can be computed from Eq.  (\ref{mbh}). Our model cannot distinguish
between the two cases, and our mass function includes both contributions. In
order to disentangle the two one would need to follow the detailed merger
history of individual haloes rather than simply using the halo mass function
at a given redshift as an input (an approach we pursue in Volonteri, Lodato \&
Natarajan, in preparation).

Massive black holes may also be related to the recent discovery of
ultra-luminous x-ray sources (ULXs) in nearby galactic discs
\citep{mush04,miller04,dewangan06}. The observed ULX luminosities of $\sim
10^{39}\,{\rm erg\,s^{-1}}$, are inconsistent with stellar mass black holes
unless they are highly beamed \citep{king01,begelman06b}. Intermediate mass
black holes (IMBHs) in the mass range $10^2\,M_{\odot}\,< M_{\rm
  BH}\,<\,10^4\,M_{\odot}$, have the appropriate Eddington luminosities to
explain ULXs. While stellar collapse is unlikely to produce this mass range of
black holes \citep{fryer01}, our proposed mechanism provides a natural way to
do so.  Recent detailed numerical studies suggest that these IMBHs might be
critical to the build-up of supermassive black holes
\citep{micic06,fregeau06}.  A fraction of IMBHs are expected to be wandering
in the haloes of galaxies as a result of frequent merging. If any of these
encounter gas-rich regions they are likely to accrete and be detected as
ULX's.

Accretion onto the seeds produced via the disc accretion scenario described
above at extremely high redshifts could power mini-quasars \citep{madau01}
that likely play a significant role in the reionization of the Universe. These
sources provide an attractive way to partially reionize the low density
inter-galactic medium, as the harder radiation emitted from these mini-quasars
in the X-ray is more likely to escape compared to the softer UV radiation from
massive stars.  Miniquasars are $\sim 10 - 100$ times more efficient at
producing ionizing radiation per processed baryon than metal-free Pop-III
stars, therefore require a much smaller reservoir of cold gas to reionize the
Universe.

\section*{Acknowledgements}

We would like to thank Andrew King and Marta Volonteri for interesting
discussions.

\bibliographystyle{mn2e} 
\bibliography{lodato}

\end{document}